\documentclass[12pt,twoside]{article}
\usepackage{fleqn,espcrc1}

\usepackage{graphicx}
\usepackage[figuresright]{rotating}

\newcommand{\AmS}{{\protect\the\textfont2
  A\kern-.1667em\lower.5ex\hbox{M}\kern-.125emS}}

\newcommand{\be}{\begin{equation}}
\newcommand{\ee}{\end{equation}}
\newcommand{\bea}{\begin{eqnarray}}
\newcommand{\eea}{\end{eqnarray}}

\usepackage{epsfig}

\title{On the Observation of Phase Transitions in Collisions of
        Elementary Matter}

\author{K.\ Paech\address{
Institut f\"ur Theoretische Physik,
Johann Wolfgang Goethe-Universit\"at,\\
Robert Mayer-Str.\ 8--10,
D-60054 Frankfurt am Main, Germany},
M.\ Reiter$^{\rm a}$, 
A.\ Dumitru\address
{Department of Physics, Columbia University,\\ 
538 West 120th Street, New York, NY 10027, USA},
H.\ St\"ocker$^{\rm a}$, and W.\ Greiner$^{\rm a,}$\footnote{Invited speaker
at CRIS 2000, 3rd Catania Relativistic Ion Studies, Acicastello, Italy, May 
22-26, 2000}
}
       
\begin{document}

\maketitle

\begin{abstract}

We investigate the excitation function of directed flow, which can
provide a clear signature of the creation of the QGP and demonstrate that
the minimum of the directed flow does not correspond to the 
softest point of the EoS for isentropic expansion. 
A novel technique measuring the compactness is introduced to determine
the QGP transition in relativistic-heavy ion collisions:
The QGP transition will lead to higher compression and therefore
to higher compactness of the source
in coordinate space. This effect can be observed by pion
interferometry. We propose to measure the compactness of the source in the
appropriate principal axis frame of the compactness tensor in coordinate space.

\end{abstract}

\section{Motivation}

The primary goal for the investigation of heavy-ion collisions is to test
the equation of state (EoS) of hot and dense matter far off the ground state,
especially with view on possible phase transitions, e.g.\ to 
the Quark-Gluon-Plasma (QGP)~\cite{QM99}.
Indeed, collective flow phenomena 
are sensitive
indicators for thermodynamically abnormal matter \cite{oldflow}.
In the case of a first-order phase transition to a QGP, an isentropic expansion
proceeds through a stage of phase coexistence
which should lead to signatures in the observables. 
The first ideas to investigate this phenomenon occured in the 
mid-seventies~\cite{Baumgardt:1975qv}.
In 
this paper we investigate the excitation function of directed flow, as well
as the compactness in heavy-ion collisions.

\section{Model}

To investigate quantitatively the experimental observables, we perform
1-fluid and 3-fluid (3+1)-dimensional relativistic 
hydrodynamic calculations. That is,
we solve numerically the continuity equations for the energy-momentum tensor,
$\partial_\mu T^{\mu\nu}=0$, and the net baryon current,
$\partial_\mu N_B^{\mu}=0$.
Detailed discussions of (3+1)-d numerical solutions for
hydrodynamical compression and expansion can be found e.g.\
in~\cite{numH}. We shall employ two different equations of
state for $P(e,\rho)$:\\
\begin{itemize}
\item[i)] 
A relativistic mean field (RMF)
hadron fluid~\cite{Serot:1986ey} corresponding
to baryons and antibaryons interacting via exchange of massive scalar and
vector bosons, plus free thermal pions; the parameters of the Lagrangian are
fitted to the ground state of infinite nuclear matter, in particular the
nuclear saturation density, the energy per particle, 
and the incompressibility.
\item[ii)] 
The same EoS as in i) for the low density phase, but supplemented by
a Bag Model EoS with a bag constant $B^{1/4}=235$~MeV for the
quark-gluon (QGP) phase. The phase coexistence region corresponding to this
first-order transition is constructed employing the Gibbs' condition of
phase equilibrium, $P_{HG}(T,\mu_B)=P_{QG}(T,\mu_B)$, where
$T$ and $\mu_B$ denote the temperature and the baryon-chemical potential,
respectively. For example, for $\rho=0$ we find $T_C\approx170$~MeV, while
at $T=0$ phase coexistence sets in at $\rho\approx4.6\rho_0$.
A more detailed discussion of these EoS can be found in~\cite{Yaris}.
\end{itemize}
Further, we employ the three-fluid model with a dynamical local unification
procedure~\cite{3faflow}.
The three-fluid model treats the nucleons of the projectile and
target nuclei as two different fluids, since they populate different
rapidity regions in the beginning of the reaction. The same holds for the
newly produced particles around midrapidity, which are therefore collected
in the third fluid. Thus, the three-fluid model accounts for the
non-equilibrium situation during the compression stage of
heavy-ion collisions. The coupling between the projectile and target fluids
is calculated assuming free binary $NN$-collisions~\cite{2fluid}.

The unification of fluids $i$ and $j$ consists of adding their
energy-momentum tensors and net-baryon currents in the respective
cells,
\be
T_i^{\mu \nu}(x) + T_j^{\mu \nu}(x) = T^{\mu \nu}_{\rm unified}(x) 
\;\;\;\; ,\;\;\;\;\; N^\mu_i(x) + N^\mu_j(x) = N^\mu_{\rm unified}(x)
\ee
and common values for $e,\, P,\,\rho$ and $u^\mu$ are obtained   
from $T^{\mu \nu}_{\rm unified} = (e+P)\, u^\mu u^\nu - P \, g^{\mu\nu}$,
$N^\mu_{\rm unified}  = \rho\, u^\mu$, and the given EoS $P=P(e,\rho)$.
The local criterion for unification is ${(P_i+P_j)}/{P} > 90\%$.
Here, $P_{i,j}$ denotes the pressure in $T^{\mu \nu}_{i,j}$,
and $P$ the pressure in $T^{\mu \nu}_{\rm unified}$.

\section{Directed flow and softest point of the EoS}

In order to measure the EoS, i.e.\ the pressure
$P(e,\rho)$ as a function of energy density $e$ and
baryon density $\rho$ in the local rest frame of a fluid element,
the transverse momentum in the reaction plane, $p_x$, is investigated.
This quantity is proportional to the pressure created in
the hot and dense collision zone \cite{oldflow}:
\be \label{eq:px}
p_x \sim \int \int P\,{\rm d} A_\perp \, {\rm d}t \quad .
\ee
The pressure $P$ is exerted on a transverse area element $A_\perp\,$.
Directed flow has therefore been proposed as a measure for the pressure
and a possible ``softening'' of the EoS~\cite{Yaris,dirflow}.

\begin{figure}[hbt]
\hfill
\begin{minipage}[t]{.6\textwidth}
\centerline{\hbox{\epsfig{figure=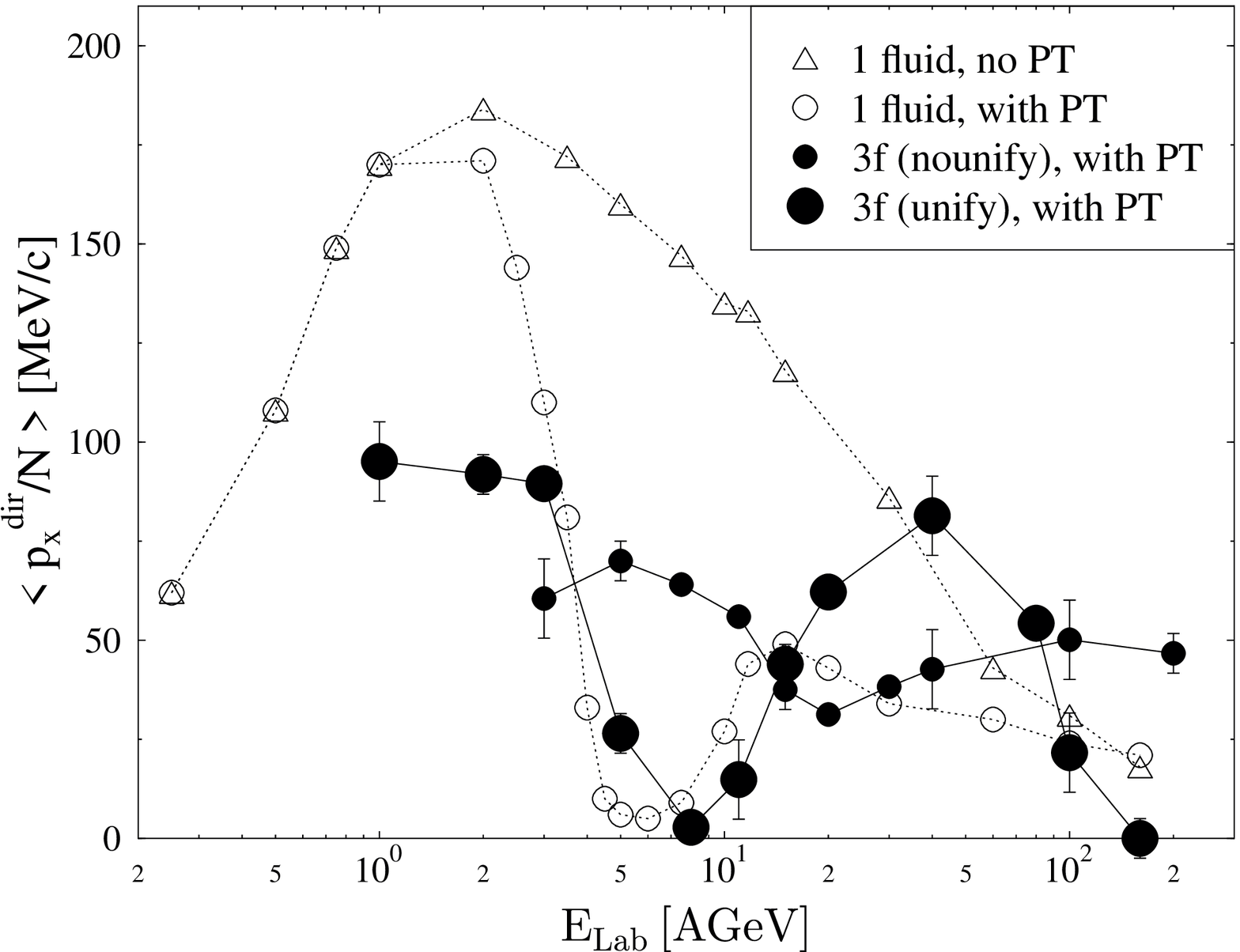,height=6cm}}}
\caption{The excitation function of directed flow $p_x^{\rm dir}/N$ for
$Au+Au$ collisions at impact para\-meter $b=3$~fm. Dotted lines
(open symbols) are
results from one-fluid dynamics; triangles are for a purely
hadronic EoS, circles are for an EoS with phase transition.
Solid lines are calculated with the three-fluid model, with (large circles)
or without (small circles) dynamical unification.
All three-fluid calculations are performed with phase transition.}
\label{pxy_3funify}
\end{minipage}
\hfill
\begin{minipage}[t]{.3\textwidth}
\centerline{\hbox{\epsfig{figure=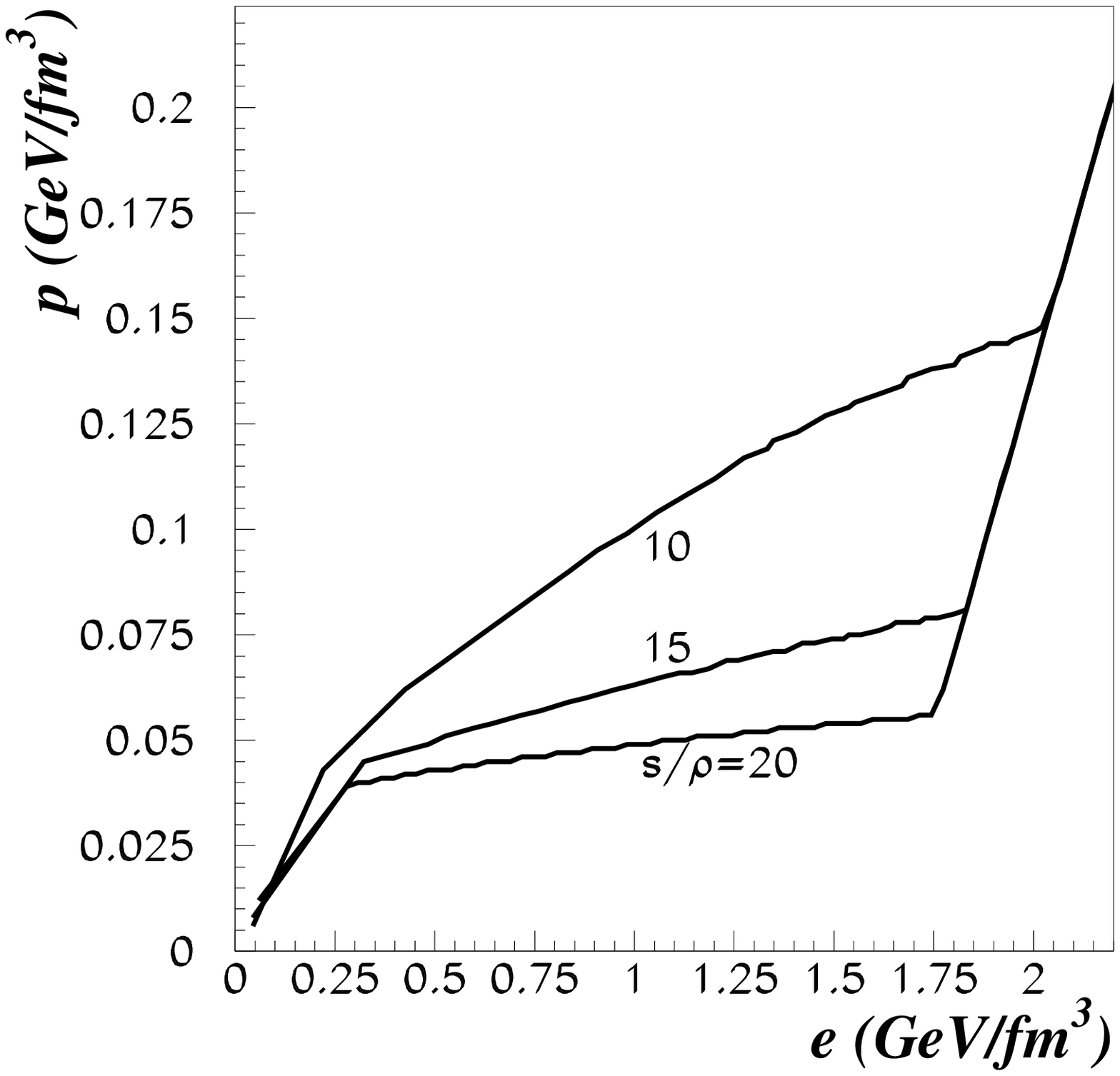,height=6cm}}}
\caption{
Pressure $P$ as a function of energy density $e$ for different entropy per
net baryon ratios for the EoS with phase transition to QGP.
}
\label{p_vs_e}
\end{minipage}
\hfill
\end{figure}

Fig.\ \ref{pxy_3funify} shows the excitation function of directed flow
$p_x^{\rm dir}/N$ calculated in the three-fluid model in comparison
to that obtained in a one-fluid calculation \cite{Yaris}.
The one fluid calculations show that
for increasing bombarding energy, the flow, $\sim p_x$, first increases (for a
collision without phase transition), as the compression and thus 
the pressure grow. At large $E_{\rm Lab}^{\rm kin}$ the time span
of the collision decreases, diminishing the flow again. The
flow is thus maximized at some intermediate bombarding energy.
In the case with a phase transition the decrease of the flow
is much more rapid than for the purely hadronic fluid.
The reason for this is not that $c_s$, the isentropic velocity of sound,
vanishes
but rather geometry: The compactness and tilt-angle $\Theta$ are different
in the calculation with phase transition, and this leads to a different
initial condition for the subsequent expansion (see below and~\cite{joergcs}).
After passing
through a local minimum at $E_{\rm Lab}^{\rm kin} \simeq5A$~GeV, the directed
in-plane momentum reaches a second local maximum around
$E_{\rm Lab}^{\rm kin} \simeq10-20A$~GeV. This is the point where the
compressed matter first becomes hot enough (over a large volume) to ``respond''
with small pressure gradients.

Due to non-equilibrium effects in the early stage of the reaction,
which delay the build-up of transverse pressure~\cite{Sorge},
the flow in the three-fluid model is reduced as compared to the
one-fluid calculation in the AGS energy range. Furthermore, the
minimum in the excitation function of the directed flow
shifts to higher energies.
The case without dynamical unification yields
the least amount of stopping and energy deposition, while
the one-fluid calculation has instantaneous
full stopping and maximum energy deposition.
The three-fluid model with dynamical unification lies between these
two limits; it accounts for the limited stopping power of nuclear matter
in the early stages of the collision and mutual equilibration
of the different fluids in the later stages.
Most importantly, the three-fluid calculations predict an
increase of $p_x^{\rm dir}/N$ towards $E_{\rm Lab}^{\rm kin}\simeq40A$~GeV,
if indeed a phase coexistence with small $c_s$ occurs. Data at that
energy has recently been taken, and should prove
very useful to pin down the onset (or absence) of a first-order
phase transition in the AGS-SPS energy domain.

First order phase transitions ``soften'' the EoS \cite{Hung,Yaris}, i.e.\
$P$ increases slower with $e$ and $\rho$ than in the case without
phase transition. This corresponds to a reduction of the isentropic
speed of sound, $c_s$, as compared to that in the interacting hadron fluid.
However, as shown in Fig.~\ref{p_vs_e}, this happens only if the entropy per
net baryon in the central region is large, i.e.\ if the ratio $T/\mu_B$ is
not too small~\cite{joergcs}. In the three-fluid model that is due mainly to
the kinetic equilibration of the decelerating baryon dense projectile/target
fluids with the midrapidity fluid of secondary particles, which leads to
considerably larger $s/\rho$ than in one-fluid hydrodynamics.
However, at the energy corresponding to the minimum of the directed flow
the specific entropy is rather small, $s/\rho\le10$; i.e.\ in the AGS energy
domain matter is rather baryon dense but not very hot. Consequently,
the EoS is {\it not} soft (i.e.\ the isentropic velocity of sound is 
not small), even if mixed phase matter does occur.

\begin{figure}[hbt]
\centerline{\hbox{\epsfig{figure=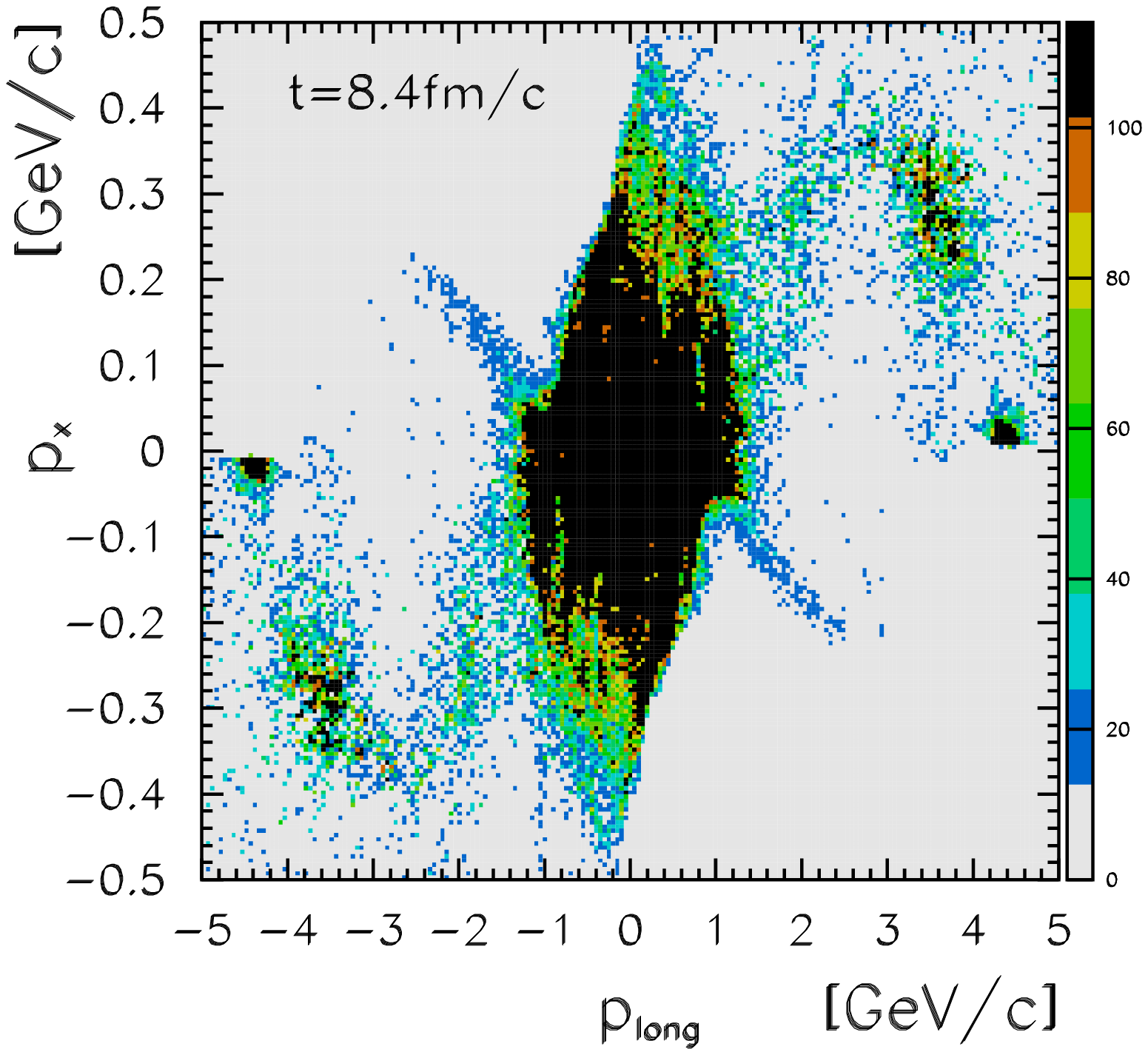,width=7.5cm}
\epsfig{figure=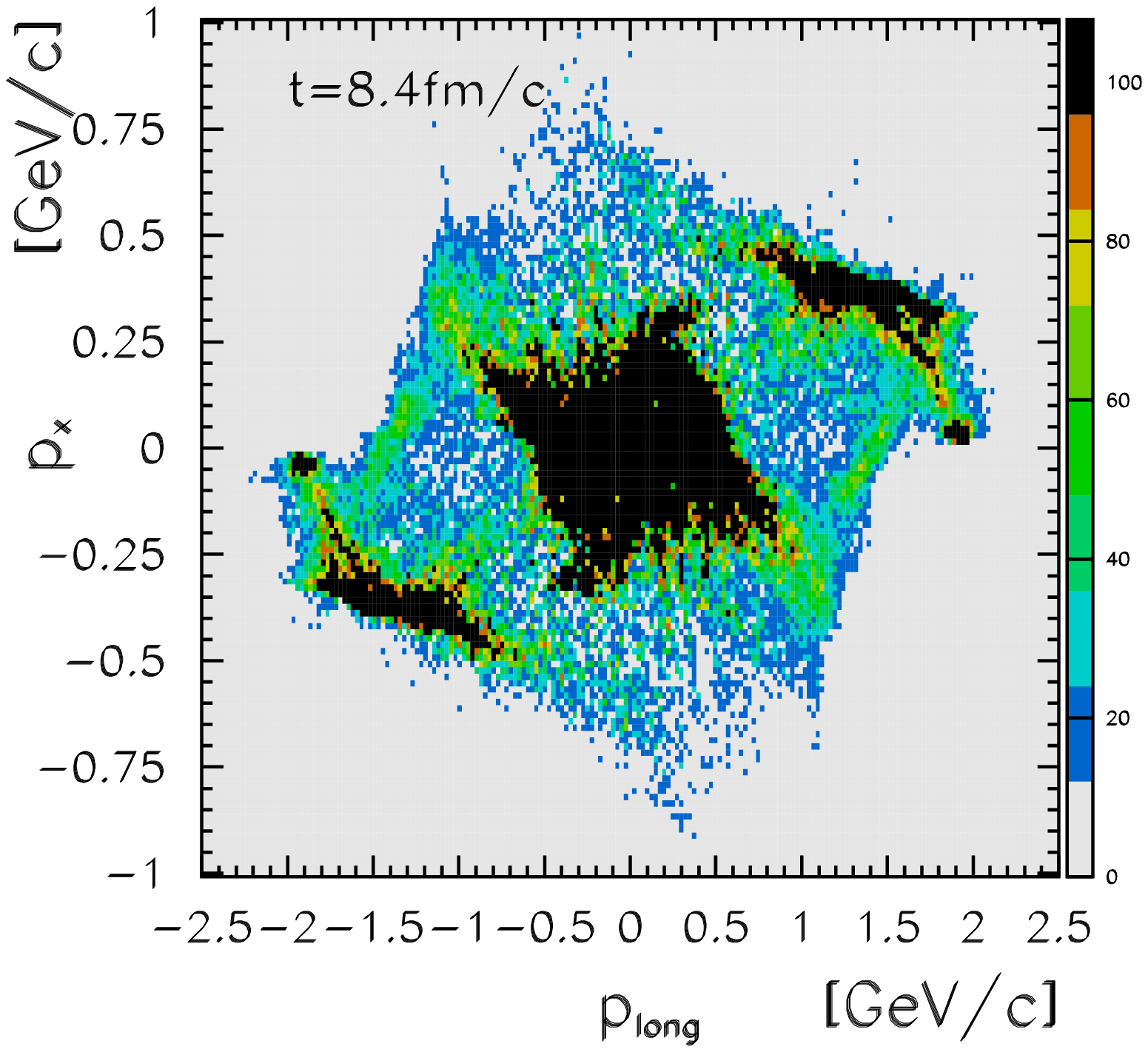,width=7.5cm}
}}
\caption{
Net-baryon density in momentum space. Pb(40~AGeV)+Pb (left; $s/\rho\approx20$)
and Pb(8~AGeV)+Pb (right; $s/\rho<10$) at $b=3$~fm.
}
\label{pxy}
\end{figure}

Fig.~\ref{pxy} shows the time-like component of the net baryon current in
momentum space ($p_x-p_{long}$ plane) for Pb+Pb-collisions at $b=3$~fm.
One clearly observes
the directed in-plane flow (before the collision, there is no matter
at $p_x\neq0$). 
However, in the left panel there is almost no momentum
of baryons in the upper left or bottom right quadrants, where
$p_x\cdot p_{long}<0$ (except for two ``jets'', see below).
The central region passed through the phase coexistence
region at high $s/\rho$, with a rather small average $c_s$, and
isentropic expansion of the highly
excited matter is inhibited. Note the
difference to the expansion pattern observed for lower
energies, right panel of Fig.~\ref{pxy},
where $\langle c_s\rangle$ is not small~\cite{joergcs}.

\begin{figure}[hbt]
\centerline{\hbox{\psfig{figure=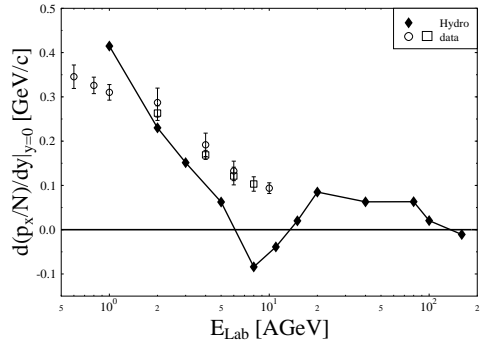,height=5.5cm}}}
\vspace*{-1cm}
\caption{
The slope of the directed in-plane momentum per nucleon at
midrapidity for Au-Au-collisions at b=3~fm (three-fluid model with dynamical
unification), the experimental data shown is from ~\cite{E895}.
}
\label{slope}
\end{figure}

The slope of $\langle p_x/N\rangle$ at midrapidity is shown in
Fig.~\ref{slope} as a function of beam energy. Experimental Data 
are shown as well~\cite{E895}.
One observes a steady decrease of $F_y={\rm d}(p_x/N)/{\rm d}y$
up to about top BNL-AGS energy, where the flow around midrapidity
even becomes negative due to preferred expansion towards
$p_x\cdot p_{long}<0$. The ``overshoot'' towards negative slope is
due to the small incompressibility in the top AGS energy region, and
the rather early fluid unification employed here.
However, such a behavior can not be observed in the data. 
A less steep decrease of $F_y$ could be achieved in the three-fluid model
by a more stringent unification criterion (i.e.\ later unification) or
early kinetic decoupling on the hadronization hypersurface.
At higher energy, $E_{Lab}\simeq40A$~GeV, we encounter the
expansion pattern depicted in the left panel of Fig.~\ref{pxy}: flow towards
$p_x\cdot p_{long}<0$ can not build up! Consequently,
$F_y$ increases rapidly towards $E_{Lab}=20-40A$~GeV,
decreasing again at even higher energy because of the more
forward-backward peaked kinematics. Note that the increase of the slope
is due to the {\em absence} of the ``anti-flow'', see Fig.~\ref{pxy}.
In any case, Fig.~\ref{slope} shows that it will be difficult to see the
effect of the possible phase transition in $F_y$. The double-differential
in-plane cross section, Fig.~\ref{pxy}, appears more useful.

\section{Compactness}

\begin{figure}[hbt]
\centerline{\hbox{\psfig{figure=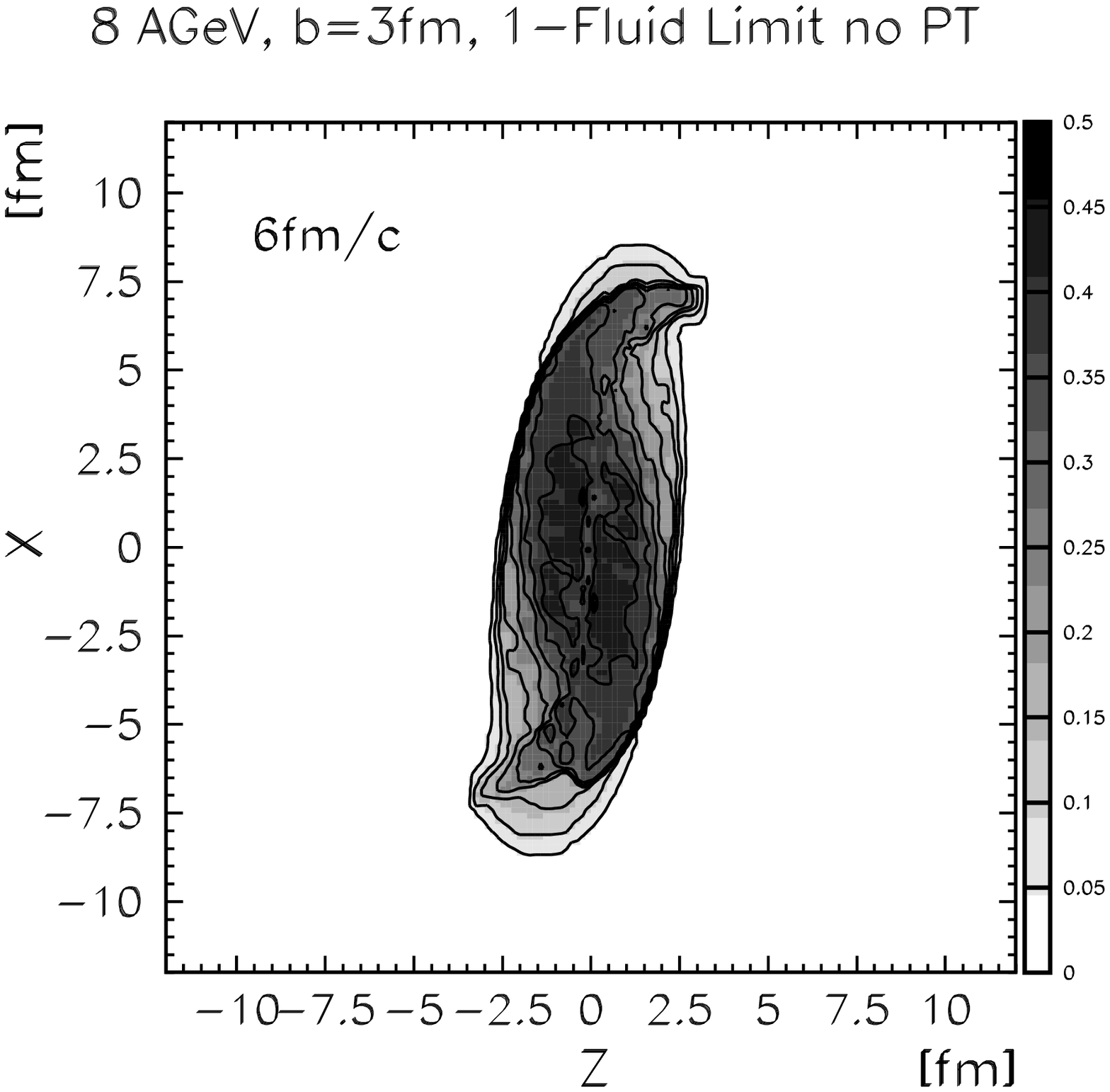,width=7.5cm}
\psfig{figure=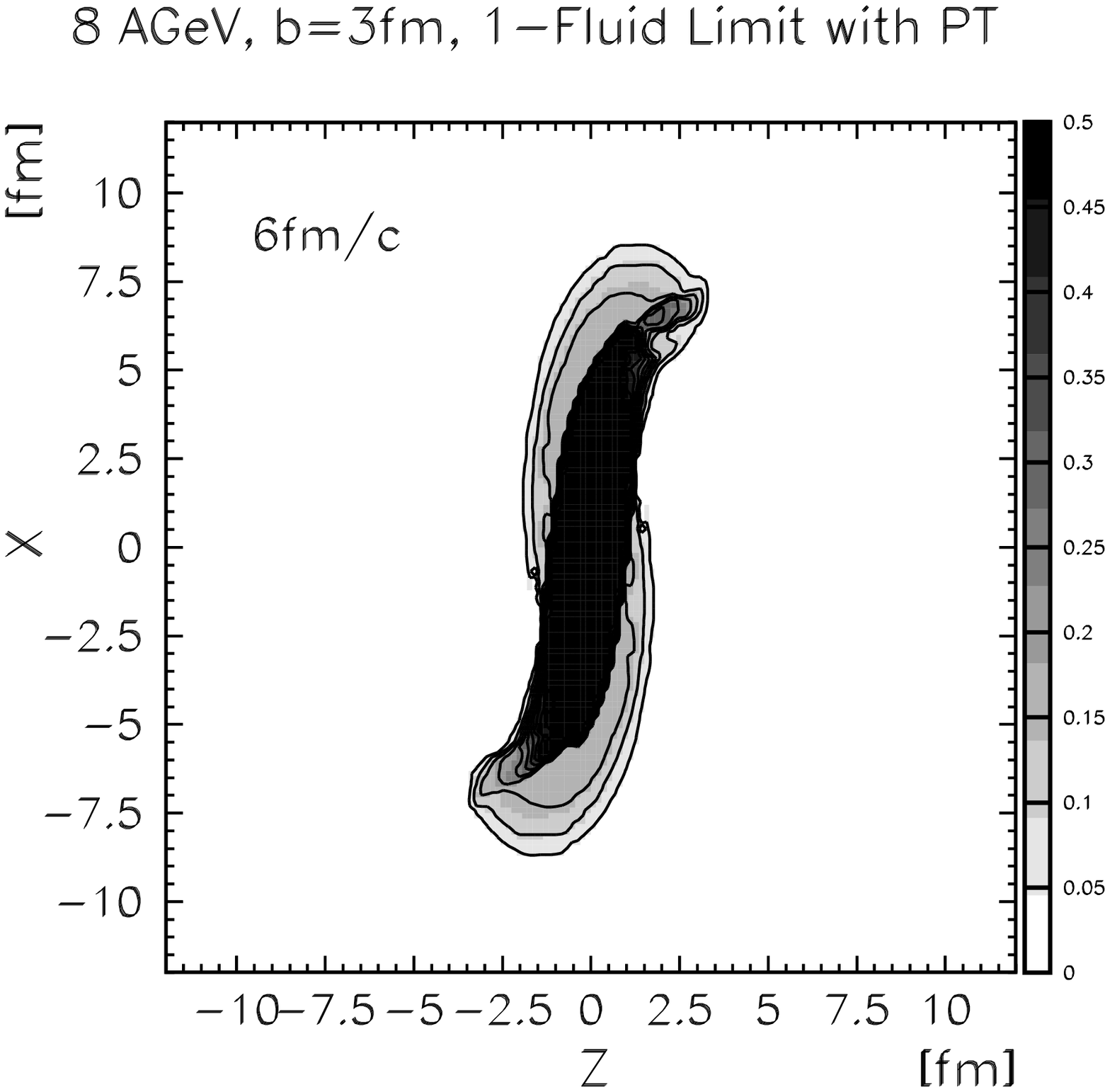,width=7.5cm}}}
\vspace*{-1cm}
\caption{
Baryon density in the reaction plane for $E_{\rm Lab}^{\rm kin}=8A$~GeV,
at time $t_{cm}=6$~fm/c.
Left: EoS without phase transition. Right: EoS with phase transition.
}
\label{comp}
\end{figure}

Measurement of the compactness is a promising new tool to observe the onset
of the phase transition. It relies
on measuring the shape of the source, which is uniquely
related to the pressure and density of the system in the compression and 
expansion stage of the nucleus-nucleus collision. The compactness can be 
identified via interferometry: The illuminiation of the baryon source by
the pion radiation is subject to experimental scrutiny via pion
interferometry~\cite{LISA}.
Fig.~\ref{comp} illustrates the basic idea.
It shows the baryon density in the reaction plane for the EoS without (i) 
and with (ii) phase transition, respectively.
One clearly observes the higher compression in the case with 
phase transition. 
As indicated above, the onset of the transition to quark matter at 
a given incident energy $E^{kin}_{lab}$ leads to
higher compression $\rho/\rho_0$ than for the
case without transition. Now, as $\rho V\simeq\pi R_A^2 L\rho$
must equal $2A$ by virtue of baryon number conservation, the longitudinal
thickness $L$ of the compressed matter is approximately 
proportional to $1/\rho$. Thus, a transition to
quark matter leads to a more compact system, just as
quark matter stars are more compact than pure neutron
stars~\cite{Glendenning:1997wn}. Of course, in
heavy-ion collisions that expectation is based on the behavior of
relativistic compression shocks rather than hydrostatic and gravitational
equilibrium.

In particular, we study the compactness in the reactions Au+Au at
impact parameter $b = 3$~fm for the energy $E^{kin}_{lab}=
8A$~GeV. The compactness is defined as the ratio of the smallest to the biggest
in-plane
eigenvalue of the configuration space sphericity tensor, which we define as
the second moment of the net baryon current. On fixed-time hypersurfaces
we have
\be
F_{ij}= \int {\rm d}^3x \,\,  x_i\, x_j N_B^0 \,\,
\Theta\left(\rho(\vec{x})-
\rho_{cut}\right)
\quad.
\ee
We apply an additional density cut $\rho>\rho_{cut}$ in the integral to discard
spectator matter. In the future the cuts and the hypersurface will have to be
adapted to the experimental conditions. However, this is not crucial for
understanding the effect.

\begin{figure}[hbt]
\centerline{\hbox{\psfig{figure=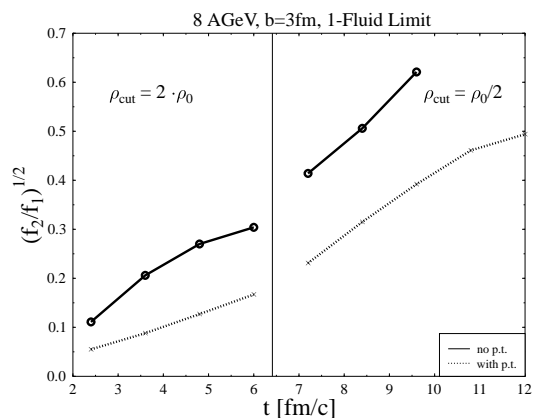,height=5.5cm}}}
\vspace*{-1cm}
\caption{Ratio of the in-plane radii $\sqrt{f_2/f_1}$
for the RMF-EoS without phase transition and for the case with transition
to quark matter.}
\label{ratio}
\end{figure}

The three eigenvalues $f_n$ are the solutions of the cubic equation
${\rm det} (F_{ij}-f\delta_{ij}) = 0$, and the eigenvectors
$\vec{e}_n$ follow from solving the
linear systems of equations $(F_{ij}-f_n\delta_{ij})e_n^j = 0$.
In terms of the eigenvalues $f_n$ and orthogonal eigenvectors
$\vec{e}_n$ of $F$, the compactness tensor can be written as
$
F = f_1 \vec{e}_1 \otimes\vec{e}_1
  + f_2 \vec{e}_2 \otimes\vec{e}_2
  + f_3 \vec{e}_3 \otimes\vec{e}_3
$.
In diagonal form, $F$ specifies an ellipsoid in configuration space with
principal axis along $\vec{e}_n$ and radii $\sqrt{f_n}$.
Cigar-like patterns, oriented along the z-axis, would lead to $f_1>f_2=f_3$,
$\vec{e}_1=\vec{e}_z$, $\vec{e}_2=\vec{e}_x$, $\vec{e}_3=\vec{e}_y$.
On the other hand, a ``pancake''/``lensil'' shape corresponds to $f_1<f_2=f_3$.
The tilt angle $\Theta$ is determined from the scalar product of
$\vec{e}_z$ (the longitudinal direction in the lab frame) with the vector
$\vec{e}_n$ corresponding to the biggest of the eigenvalues $f_n$.
As already indicated in the introduction, we find very different eigenvalues
for the two equations of state.
The calculation with transition to quark
matter corresponds to higher compactness of the baryon distribution.
That is, the compactness tensor is much flatter (nearly a factor of two~!)
in the model with (ii) than in the model without (i) phase transition.
Moreover, our (3+1)-dimensional expansion solutions show that
after the compression stage the ratio of the in-plane
radii $\sqrt{f_2/f_1}$ remains much smaller in the case with phase
transition, cf.\ Fig.~\ref{ratio}. Note also that the extremely flat shape
of the baryon distribution means very small curvature of the surface, which in
turn will lead to a strongly ``bundled'' emission of hadrons from the
(almost planar) rarefaction wave or deflagration shock by which the dense
droplet decays; see also the discussion in~\cite{Danielewicz:1994nb}.

The eigenvalues of the compactness tensor allow to measure directly the density
increase in the high density stage of the reaction, if a phase transition
occurs.
Care must be taken that the impact parameter range investigated constitutes a
moderately small bin of centrality values. One should keep in mind
that the compression factor is affected by the incompressibility
$\partial P/\partial e$ evaluated on the shock adiabat (in the
one-fluid model), {\em not} along a path of fixed specific entropy.
Therefore, the incompressibility needs not be equal to the isentropic
speed of sound. The latter is not much reduced in the presence of the phase
transition at AGS energies~\cite{joergcs}, because of the high net baryon
density and relatively low temperature.

Lisa et al.~\cite{LISA} have recently proposed a new 
interferometry analysis, which
could be used to observe this change in the compactness directly.
The developed method is quite robust and incorporates other interesting
information as the configuration space tilt angle, which
nicely complement the momentum-space flow angles.
It avoids cuts in tilted ellipsoids, which are not
analysed in the appropriate rotated frame, and where the excentricity and
the RMS-radii are much less distinct for the two different equations 
of state.



\end{document}